\documentclass[aps,prl,floatfix,epsfig,twocolumn,showpacs,preprintnumbers]{revtex4}
\usepackage{amsmath,amssymb,graphicx,bm,epsfig}
\usepackage{color}
\usepackage{braket}
\usepackage{hyperref}

\renewcommand{\vr}{{\mathbf{r}}}

\newcommand{\Tr}{\mathrm{Tr}}

\newcommand{\Ang}{\textrm{\AA}}

\begin{document}

\title{Free energy from stationary implementation of the DFT+DMFT functional}

\author{Kristjan Haule}
\author{Turan Birol}
\affiliation{Department of Physics and Astronomy, Rutgers University, Piscataway, USA}
\date{today}

\begin{abstract}
  The stationary functional of the all-electron density functional plus
  dynamical mean field theory (DFT+DMFT) formalism to perform free
  energy calculations and structural relaxations is implemented for
  the first time. Here, the first order error
  in the density leads to a much smaller, second order error in the free energy.  The
  method is applied to several well known correlated materials;
  metallic SrVO$_3$, Mott insulating FeO, and elemental Cerium, to
  show that it predicts the lattice constants with very high
  accuracy.  In Cerium, we show that our method predicts the
  iso-structural transition between the $\alpha$ and $\gamma$ phases, and
  resolve the long standing controversy in the driving mechanism of this
  transiton.
\end{abstract}

\date{\today}
\maketitle

Prediction of the crystal structures of solids by large scale quantum
mechanical simulations is one of the fundamental problems of condensed
matter physics, and  occupies a central place in materials design.
The workhorse of the field is the Density Functional Theory
(DFT)~\cite{DFT} at the level of Local Density Approximation (LDA) or 
Generalized Gradient Approximations (GGAs), which predict lattice constants of weakly
correlated materials typically within $\sim$1\% relative error~\cite{Kresse}.

These errors of DFT in LDA/GGA implementations are an order of magnitude
larger in the so called correlated materials: For example, the lattice
constant of $\delta$-Pu is underestimated by 11\%~\cite{SavrasovNat}
or non-magnetic FeO by 7\%\cite{online}. While GGAs and hybrid functionals can
sometimes improve upon conventional LDA, these functionals many times
degrade the agreement between predicted and experimentally determined
bulk moduli and lattice constants, in particular in materials
containing heavy elements.~\cite{Kresse}

To account for the correlation
effects, more sophisticated many body methods have been
developed. Among them, one of the most successful algorithms is the
dynamical mean-field theory (DMFT)~\cite{GabiFirst}. It replaces the
problem of describing correlation effects in a periodic lattice by a
strongly interacting impurity coupled to a self-consistent
bath~\cite{rmp}. To become material specific, DMFT was soon developed
into an electronic structure tool
(LDA+DMFT)~\cite{KotliarFirst,Lichtenstein}, which achieved great
success in numerous correlated materials (for a review
see~\cite{rmp2}). The LDA+DMFT method has mainly been used for the
calculation of spectroscopic quantities, and only a few
dozens~\cite{savrasov2001,mcmahan2001,mcmahan2003,amadon2006,Savrasov_prl06,
georges2007,vollhardt2008,lichtenstein2009,georges2011,haule2012,amadon2012,vollhardt2012,eriksson2012,wills2012,lechermann2012,pourovskii2013,vollhardt2014,amadon2014,cohen2014,park2014a,park2014b} 
of studies managed to compute energetics of
correlated solids, and only a handful of them used exact solvers and
charge self-consistency~\cite{cohen2014, georges2011,haule2012,lechermann2012,pourovskii2013,park2014a}. 
This is not only because of
the very high computational cost, but also because previous
implementations of LDA+DMFT were not stationary, and hence it was hard
to achieve precision of free energies needed for structure
optimization and study of phase transitions in solids.

Here we implemented the LDA+DMFT functional, which delivers stationary
free energies at finite temperatures. This stationarity is crucial for
practical implementation and precision of computed energies, since the
first order error in the density $\rho$ (or the Green's function)
leads only to the much smaller second order error in the free energy,
since the first order variation vanishes, i.e., $\delta
F/\delta\rho=0$. This property is also crucial in calculating the
forces, as stationarity of the functional ensures that only
Hellmann-Feynman forces need to be computed for structural relaxation
.~\footnote{Note that Pulay forces appear when incomplete basis set is used.}

The DFT+DMFT total energy is given by~\cite{rmp2} :
\begin{eqnarray}
E = \Tr(H_0 G) + \frac{1}{2}\Tr(\Sigma G) + E^H[\rho]+E^{xc}[\rho]
\nonumber\\
- \Phi^{DC}[n_{loc}] +E_{nuc-nuc}
\label{Eq:Ene}
\end{eqnarray}
where $H_0 = -\nabla^2 + \delta(\vr-\vr')V_{ext}(\vr)$, 
$G$ is the electron Green's function, $E^H[\rho]$ and $E^{xc}[\rho]$ are Hartree and
DFT exchange-correlation functional, $V_{ext}$ is the electron-nuclear
potential, $E_{nuc-nuc}$ is the interaction energy of nuclei, $\Sigma$
is the DMFT self-energy, and $\Phi^{DC}[n_{loc}]$ is the
double-counting (DC) functional.~\cite{online} Here the
Migdal-Galitskii formula (MGF) is used $E_{pot}=\frac{1}{2}\Tr(\Sigma
G)$ to compute the DMFT part of the potential energy.

Gordon Baym showed~\cite{Baym} that for certain class of
approximations, which are derivable from a functional expressed in
terms of closed-loop Feynman diagrams, MGF can be used instead of more
complicated expression for evaluating the Luttinger-Ward
Functional~\cite{Luttinger-Ward,Baym-Kadanoff}. He called such
approximations conserving. While the DMFT is a conserving approximation
in Baym's sense, LDA or GGA are not, as the Galitskii-Migdal formula
$\frac{1}{2}\Tr(V_{xc}\rho)$ has to be replaced by the
exchange-correlation functional $E^{xc}[\rho]$.  As a result, the
combination of DFT+DMFT in its charge-self consistent version is not
conserving either, and consequently MGF can give different
total energy than the Luttinger-Ward functional.  Only the evaluation
of the latter is guaranteed to give stationary free energies. We will
give numerical evidence that evaluation of MGF in Eq.~\ref{Eq:Ene}
gives different results than evaluation of the Luttinger-Ward
functional, which strongly suggests that Eq.~\ref{Eq:Ene} gives
non-stationary total energies.

The Luttinger-Ward functional of DFT+DMFT has been well known for several
years~\cite{rmp2}, but it has never been successfully implemented to
%
%
%
compute the free energy of a solids. It has the following form
\begin{eqnarray}
\Gamma[G] = \Tr\log G-\Tr((G_0^{-1}-G^{-1})G) +
E^{H}[\rho] \nonumber\\
+E^{xc}[\rho]+\Phi^{DMFT}[\hat{P}G]-\Phi^{DC}[\hat{P}\rho]+E_{nuc-nuc},
\label{Eq:Funct}
\end{eqnarray}
where
$G_0^{-1}(\vr\vr';i\omega)=[i\omega+\mu+\nabla^2-V_{ext}(\vr)]\delta(\vr-\vr')$,
$\Phi^{DMFT}[\hat{P} G]$ is the DMFT functional, which is the sum of all local skeleton
Feynman diagrams.
The projected Green's function $\hat{P}G\equiv
G_{local}=\sum_{LL'}\ket{\phi_L}\braket{\phi_L|G|\phi_{L'}}\bra{\phi_{L'}}$
and the projected density
$\hat{P}\rho\equiv\rho_{local}$ are computed with projection 
to a set of localized functions $\ket{\phi}$ centered on
the "correlated" atom. The projection defines the local Green's
function $G_{local}$,
the essential variable of the DMFT.

The variation of functional $\Gamma[G]$ with respect to $G$
($\delta\Gamma[G]/\delta G$) gives, 
\begin{eqnarray}
G^{-1}-G_0^{-1} + (V_H + V_{xc})\delta(\vr-\vr')\delta(\tau-\tau')\nonumber\\
+\hat{P}\frac{\delta \Phi^{DMFT}[G_{local}]}{\delta G_{local}}
\nonumber\\
-\hat{P}\frac{\delta \Phi^{DC}[\rho_{local}]}{\delta \rho_{local}}\delta(\vr-\vr')\delta(\tau-\tau')=0,
\label{Dyson}
\end{eqnarray}
which vanishes, since it is equal to the Dyson equation that
determines self-consistent $G$, hence the functional is stationary.  


The value of the functional $\Gamma$
at the self-consistently determined $G$ delivers the free energy of the
system~\cite{Baym}. We evaluate it by inserting $G_0^{-1}-G^{-1}$ from
Eq.~\ref{Dyson} into Eq.~\ref{Eq:Funct} to obtain
\begin{eqnarray}
F  = E_{nuc-nuc}
-\Tr((V_H+V_{xc})\rho)
+E^{H}[\rho]+E^{xc}[\rho]\nonumber\\
+\Tr\log G
-\Tr\log G_{loc}+F_{imp}\nonumber\\
+\Tr(V_{dc}\rho_{loc}) 
-\Phi^{DC}[\rho_{loc}]+\mu N,
\label{Eq:F}
\end{eqnarray}
where we denoted
$V_{dc}\equiv \delta \Phi^{DC}[\rho_{local}]/\delta \rho_{local}$
and $F_{imp}$ is the free energy of the impurity problem, i.e.,
$F_{imp}=\Tr\log G_{loc}-\Tr(\Sigma G_{loc})+\Phi^{DMFT}[G_{loc}]$.~\cite{online}
Here we also use the fact the solution of the auxiliary impurity
problem delivers the exact local Green's function, i.e.,
$\Sigma={\delta \Phi^{DMFT}[G_{local}]}/{\delta G_{local}}$, and
we added $\mu N$ because we work at constant electron number.

The crucial point is that the continuous time quantum Monte Carlo
method (CTQMC)~\cite{hauleCTQMC,Werner} solves the quantum impurity
model (QIM) numerically exactly, hence, we can compute very precisely
the impurity internal energy as well as the free energy $F_{imp}$ of
this model. At high enough temperature, $F_{imp}$ can be directly
read-off from the probability for the perturbation order $k$, which we
call $P_k$, and is computed by $P_k=Z_k/Z$ (where $Z_k$ is the
partition function with $k-$kinks), hence $Z=Z_{atom}/P_0$, where
$Z_{atom}$ can be directly computed from atomic energies, and $P_0$ is
the probability for no kinks on the impurity. Finally,
$Z=\exp(-F_{imp}/T)$, giving directly
\begin{equation}
F_{imp}=-T (\log(Z_{atom})-\log(P_0)).
\label{Fimp}
\end{equation}

When the temperature is low, $P_0$
becomes exponentially small, and we can no longer determine $Z$ to high
enough precision in this way. However, we can compute very precisely
the internal energy of the impurity at arbitrary temperature. The
internal energy of QIM $E_{imp}$ is given by
\begin{eqnarray}
E_{imp}=\Tr((\Delta+\varepsilon_{imp}-\omega_n\frac{d\Delta}{d\omega_n})
G_{imp}) + E_{imp-pot},
\end{eqnarray}
which follows directly from the thermodynamic average of QIM
Hamiltonian. Here the hybridization $\Delta$ and impurity levels
$\varepsilon_{imp}$ are determined from the local green's function by
the standard DMFT self-consistency condition
$G_{local}^{-1}=i\omega_n-\varepsilon_{imp}-\Sigma-\Delta$, and
$E_{imp-pot}=\frac{1}{2}\Tr(\Sigma G_{imp})$.  These quantities
can be computed very precisely by CTQMC using the following tricks: i)
$\Tr(\Delta G_{imp})$ is computed from the average perturbation order
$\braket{k}$ of CTQMC, and takes the form $\Tr(\Delta
G_{imp})=\braket{k}/T$, where $T$ is temperature~\cite{hauleCTQMC}; ii)
$E_{imp-pot}$ is computed from the energies of atomic state of QIM
$E^{atom}_m$ and their probabilities $P_m$ by $E_{imp-pot}=\sum_m P_m
E^{atom}_m - \Tr(\varepsilon_{imp} n_{imp})$~\cite{hauleCTQMC}, which delivers
much more precise interaction energy than obtained by MGF; ii) We
spline $\Delta(\omega_n)$ in Matsubara points and determine its
derivative $d\Delta/d\omega_n$, and then carry out Matsubara sum by
subtracting out the leading high-frequency tails by formula
$A/((i\omega-\varepsilon_1)(i\omega-\varepsilon_2))$, which has an
analytic sum of
$A(f(\varepsilon_1)-f(\varepsilon_2))/(\varepsilon_1-\varepsilon_2)$.
Because probabilities $P_m$ and perturbation order $\braket{k}$ are
known to very high precision in CTQMC, the impurity internal energy
can easily be computed with precision of a fraction of a meV.

To compute precise impurity free energy $F_{imp}$ at lower temperature, we first
converge DFT+DMFT equations to high accuracy at low temperature. Using converged impurity
hybridization $\Delta(i\omega_n)$, 
%
%
we raise the temperature of the impurity (keeping $\Delta$ fixed) to $T_>$, so that $P_0$
is of the order of $10^{-5}$ or higher, and obtain reliable $F_{imp}$
using Eq.~\ref{Fimp}, and entropy $S_{imp}$ at this higher temperature
$S_>=(E_{imp}(T_>)-F_{imp}(T_>))/T_>$.
Next, we evalute impurity internal energy for several inverse temperatures
$\beta=1/T$, 
and than we use standard thermodynamic relations to obtain entropy at
lower temperature $T$ by
\begin{eqnarray}
S(T)=S_>-\frac{E_{imp}(T_>)}{T_>} +
\frac{E_{imp}(T)}{T}-\int_{1/T_>}^{1/T} E_{imp}(\beta)d\beta
\label{Eq:S}
\end{eqnarray}
where $\beta=1/T$. This formula is obtained integrating by parts
the standard formula $S=\int c_v/T dT$ and $c_v=dE/dT$. We hence
obtain $S_{imp}$ and  $F_{imp}=E_{imp}-T S_{imp}$ at lower $T$ which can be inserted
into Eq.~\ref{Eq:F}.
The rest of the terms in Eq.~\ref{Eq:F} are relatively straightforward to
evaluate, however, for a high precision implementation one needs to
combine the terms that largely cancel and evaluate them
together~\cite{online}.

Previous implementations of free energy within
LDA+DMFT~\cite{held2001,pourovskii2013,amadon2014} were based on i)
evaluating the total energy Eq.~\ref{Eq:Ene} at range of temperatures, and
integrating resulting specific heat~\cite{held2001}, and ii) the
coupling constant integration~\cite{pourovskii2013,amadon2014}, where
total energy of the solid is needed for a range of coulomb repulsion's
$U$ and is than integrated over $U$. In both approaches, the
self-consistent LDA+DMFT solution is needed for many values of the
parameters (either U or T) to evaluate $F$.  In our method, a single
LDA+DMFT calculation for solid is needed, which makes the method much
more efficient. 
Furthermore, current implementation of the free energy is stationary,
hence higher precision of $F$ is achieved.

To test the implementation of the LDA+DMFT functional, we computed the
volume dependence of the free energy for three well studied correlated
materials: a metallic early transition metal oxide with perovskite
structure SrVO$_3$, a Mott insulating transition metal oxide FeO in
its rock salt structure, and the lanthanide elemental metal, Cerium,
in its face centered cubic structure, which undergoes a first order iso-structural
transition.

We used the implementation of LDA+DMFT of
Ref.\onlinecite{hauleLDADMFT}, which is based on the Wien2K
package~\cite{wien2k}, and LDA in combination with nominal
double-counting~\cite{hauleLDADMFT,covalency,exactDC}. More technical
details are given in the supplementary information.

\begin{figure}
\includegraphics[width=0.93\linewidth]{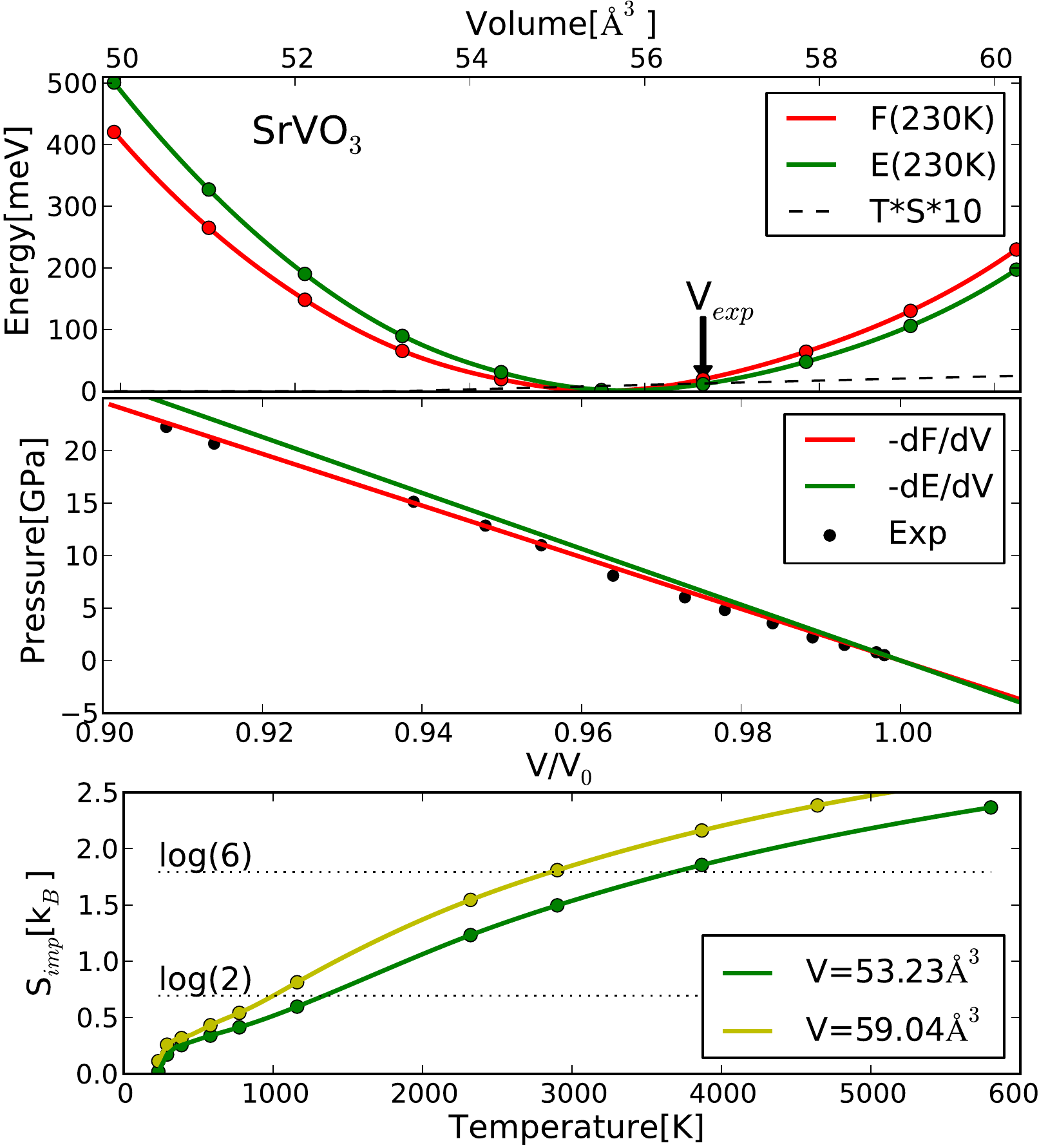}
\caption{ (Color online): a) E(V) and F(V) for SrVO$_3$ at $T=230\,$K
  from Eq.~\ref{Eq:Ene} and ~\ref{Eq:F}, respectively. Entropy term $T
  S_{imp}(V)$ is very small. (b) theoretical and
  experimental~\cite{SrVO3:pv} $p(V)$. Good agreement between
  theoretical $-dF/dV$ and experiment is found. (c) Impurity Entropy
  Eq.~\ref{Eq:S} for representative volumes. To obtain $S_{imp}$,
  temperature is varied in the impurity problem only, and not in the
  LDA+DMFT problem of the solid.  }
\label{fig:srvo}
\end{figure}
In Fig.~\ref{fig:srvo}(a) we show the energy $E(V)$, and $F(V)$ for
SrVO$_3$ at $T=230\,$K, computed with Eq.~\ref{Eq:Ene}, and
Eq.~\ref{Eq:F}, respectively. The minima of $E(V)$ and $F(E)$ are
achieved at $55.71\,\Ang^3$ and $55.51\,\Ang^3$. The experimentally
determined volume is $V_{exp}=56.53\,\Ang^3$~\cite{SrVO3:vol}.
The LDA+DMFT hence slightly underestimates the equilibrium volume ($1.8\%$),
which gives $0.6\%$ error in lattice constant. This is well within the
standard error of best DFT functionals for weakly correlated
materials.

The metallic nature of SrVO$_3$, with moderate mass enhancements
$m^*/m_{band}\approx 2$~\cite{online}, leads to very small DMFT
corrections in crystal structure~\cite{online}.  
%
%
Note that energy minimization leads to slightly larger volume
than free energy minimization, contrary to expectations. This is
because energy is computed from non-stationary Eq.~\ref{Eq:Ene}, while
free-energy is obtained from the stationary expression
Eq.~\ref{Eq:F}. The latter is hence more trustworthy, and should be
considered best LDA+DMFT result. This is also clear from pressure
versus volume diagram in Fig.~\ref{fig:srvo}(b), where $-dF/dV$
agrees more favorably with the experiment than $-dE/dV$ obtained by MGF.

In Fig.~\ref{fig:srvo}(c), we show the impurity entropy obtained by
Eq.~\ref{Eq:S} for two representative volumes. In this itinerant
system with very large hybridization, we do not notice a temperature
scale at which $t2g$ shell is degenerate ($log(6)$) nor the scale of
the lowest order Kramers doublet ($log(2)$), but we notice the Fermi liquid
scale in the steep downturn of $S(T)$ at $T\approx 350\,$K.

\begin{figure}
\includegraphics[width=0.86\linewidth]{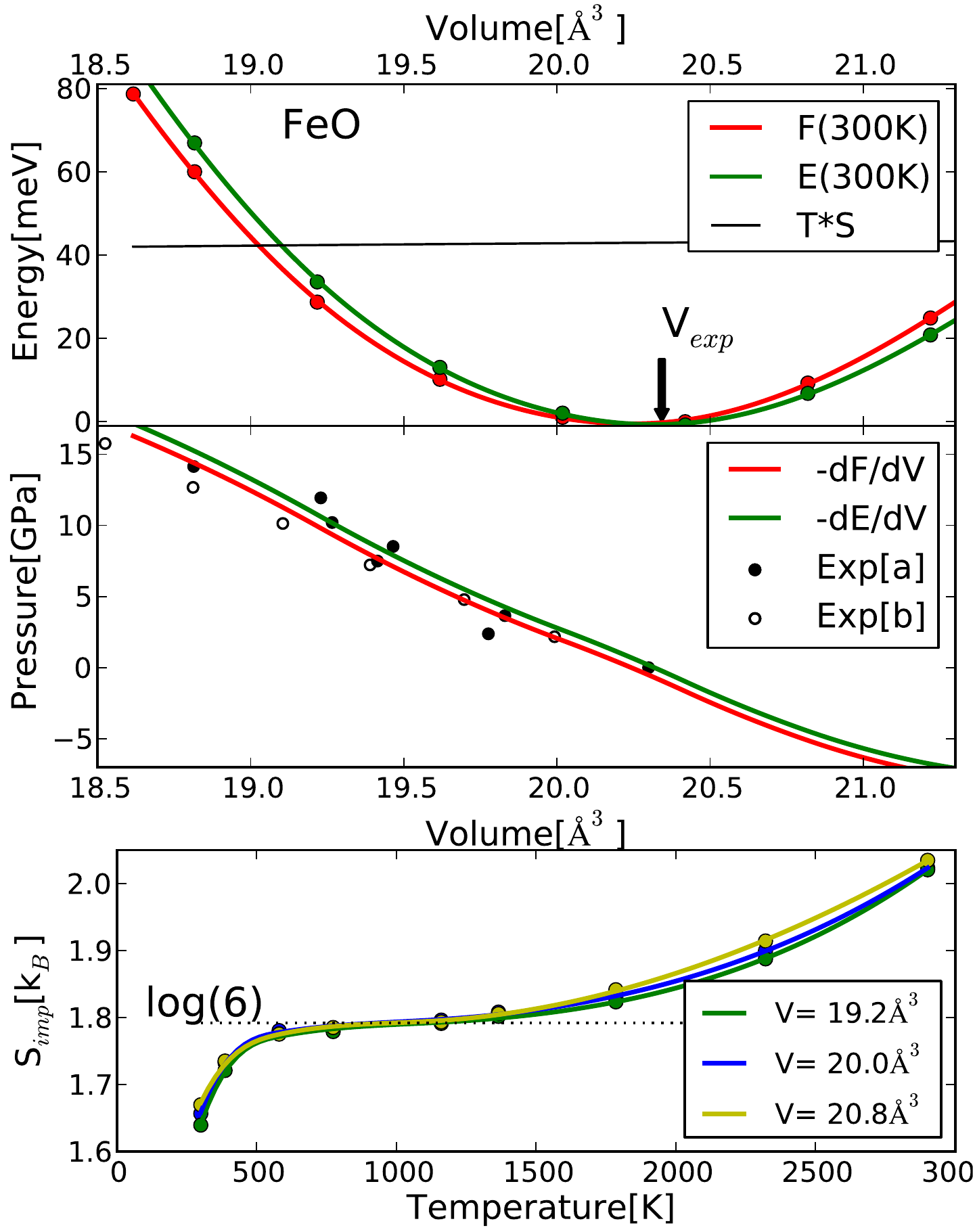}
\caption{
(Color online): a) E(V) and F(V) for FeO from Eq.~\ref{Eq:Ene} and
~\ref{Eq:F}, respectively. Entropy term $T S_{imp}(V)$ is large but
almost constant. (b) theoretical and experimental
$p(V)$. Filled and empty circles are from
Refs.~\onlinecite{FeO:filled} and~\onlinecite{FeO:empty}, respectively. 
(c) Impurity entropy Eq.~\ref{Eq:S} for representative
volumes. The degeneracy of the $t_{2g}$ shell above 1000K is apparent.
}
\label{fig:feo}
\end{figure}
Fig.~\ref{fig:feo}(a) shows $E(V)$ and $F(V)$ for paramagnetic Mott
insulating FeO at $300\,$K, above its antiferromagnetic ordering
temperature. The equilibrium volume of $E$ and $F$ is $20.28\Ang^3$
and $20.24\Ang^3$, while the experimental volume is
$20.342\Ang^3$. The lattice constant is thus underestimated for only
$0.10\%$ and $0.16\%$, respectively. In comparison, all standard DFT functionals
severally underestimate FeO lattice constant, for example PBE-sol,
PBE, and LDA for $5.2\%$, $5.0\%$, and  $7.7\%$, respectively. 

In Fig.~\ref{fig:feo}(b) we show $P(V)$ diagram and its excellent
agreement with experiment. Fig.~\ref{fig:feo}(c) shows impurity
entropy $S_{imp}(T)$ for a few volumes. In contrast to metallic
SrVO$_3$, here we clearly see an extended
plateau of $S_{imp}(T)=log(6)*k_B$ around $1000\,$K, which signals
complete degeneracy of the $t_{2g}$ shell, and its slight decrease at
$300\,K$ in proximity to the AFM state.

\begin{figure}
\includegraphics[width=0.86\linewidth]{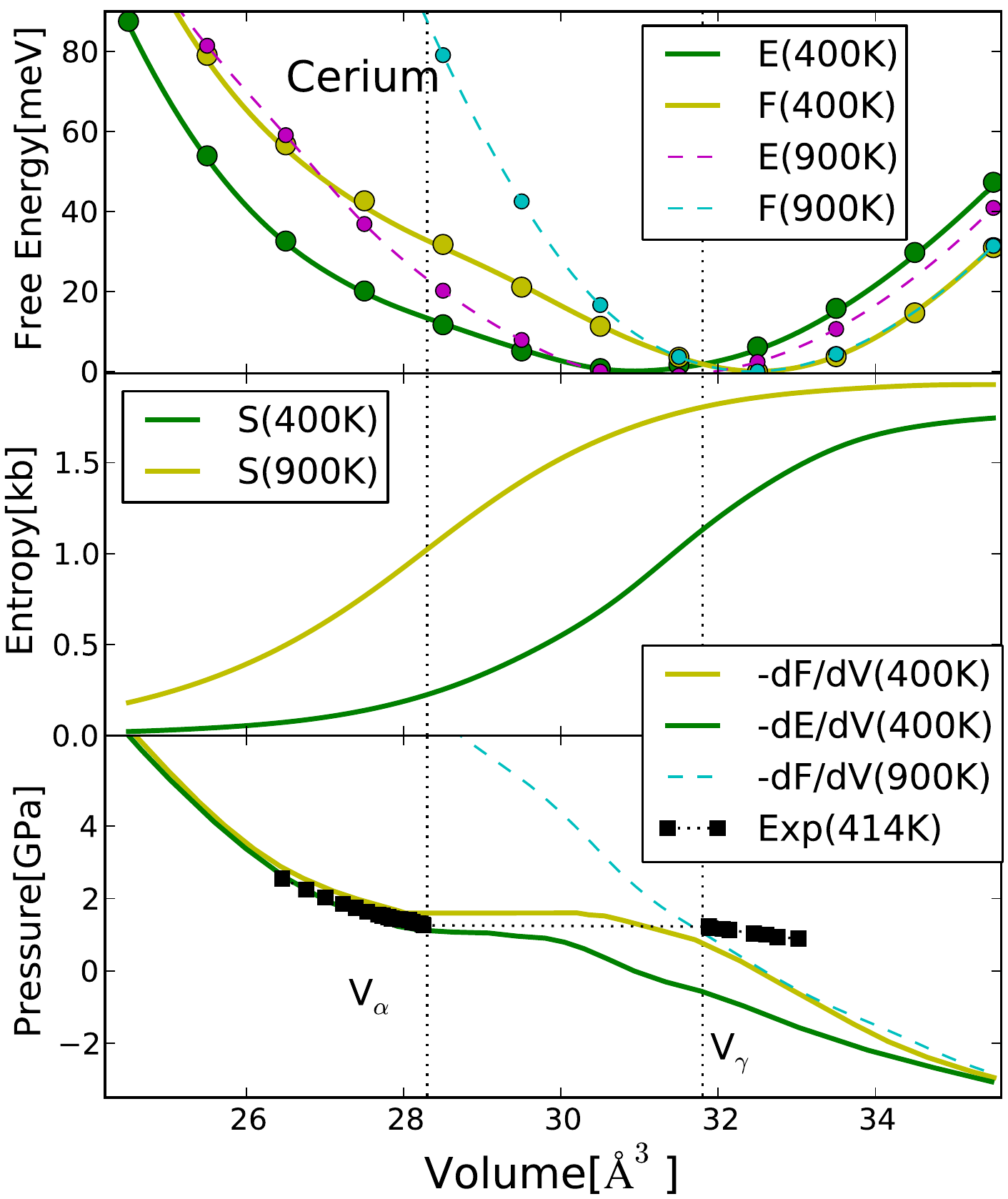}
\caption{ (Color online): a) E(V) and F(V) for elemental Cerium from
  Eq.~\ref{Eq:Ene} and ~\ref{Eq:F}, respectively. Data are presented
  for T=400 and 900K. (b) Entropy $S_{imp}(V)$ is large and changes
  dramatically accros the transiton.
(c) theoretical and experimental ~\cite{Ce:exp} $p(V)$ diagram.
}
\label{fig:ce}
\end{figure}
 The iso-structural transitions of Cerium attracted a lot of experimental and
theoretical effort, but its theoretical understanding is still
controversial. On the basis of LDA+DMFT calculation
McMahan~\textit{et.al}~\cite{mcmahan2001} proposed that the total
energy exhibits a double-minimum shape, concomitant with the
appearance of the quasiparticle peak at temperature as high as
$1500\,$K, signaling the first order transition. Using different
implementation of the same method,
Amadon~\textit{et.al}~\cite{amadon,amadon2014} proposed that the
transition is entropy driven, and that the total energy is featureless
with the minimum corresponding to low volume $\alpha$-phase. Only the
addition of the entropy term moves the minimum to the larger volume of
$\gamma$-phase. 
In this picture the transition at low temperatures, where the entropy 
becomes small and cannot drive the transition, is intrinsically absent. 
%
%
Yet another proposal was recently put forward on
the basis of LDA+Gutzwiller calculations~\cite{Lanata:2013,Lanata:2014}, in
which the transition is present even at zero temperature, but the
transition occurs at negative pressure. The transition is thus
detectable even in the total energy, in the absence of entropy, and
becomes second order at $T=0$. In the same method, the finite
temperature transition is first order, and the double-minimum shape of
free energy becomes most pronounced at very high temperature
($1500\,$K)~\cite{Lanata:2014}.

Our LDA+DMFT results for Ce are plotted in Fig.~\ref{fig:ce}. The
total energy curve at $400\,$K clearly shows a region of very flat
shape in the region between the $\alpha$-$\gamma$ volume. Indeed the
derivative of the energy $-dE/dV$ displayed in Fig.~\ref{fig:ce}(c)
shows a clear region of zero slope around $1\,$GPa. This is consistent
with results of Lanata \textit{et al.}~\cite{Lanata:2013} finding very
similar zero slope of $-dE/dV$ at zero temperature, but is
inconsistent with Ref.~\onlinecite{amadon2014}, which finds no feature
in total energy. It is also inconsistent with
McMahan~\textit{et.al}~\cite{mcmahan2001} showing clear double-peak in
total energy. On the other hand, the addition of entropy substantially
increase the region of soft volume, as suggested by
Amadon~\textit{et.al}~\cite{amadon}. Indeed the change of the entropy
between the two phases is of the order of 0.9$k_B$, which is
consistent with experimental estimations of $30\,$meV at
400K~\cite{expS}. The physical mechanism behind this large entropy
change and unusual volume dependence of energy is in very fast
variation of coherence temperature, as suggested in
Refs.~\cite{mcmahan2001,amadon}, and conjectured in Kondo volume
collapse theory~\cite{KVC}.  The phase transition in our calculation
occurs around $1.6\,$GPa, which is not far from experimentally
determined critical pressure of $1.25\,$GPa at $T=400\,$K. The free
energy barrier in our calculation is however extremely small, and no
clear double peak of $F(V)$ or negative slope of $-dF/dV$ can be
detected within our $1\,$meV precision of energies. This is similar to
results of Ref.~\onlinecite{Lanata:2014} at $400\,$K, but different
from Ref.~\onlinecite{mcmahan2001}. While the start of the transition
region in $\alpha$-phase is in good agreement with experiment, the
$\gamma$-phase volume is underestimated in our calculation. We believe
that the addition of phonon entropy is needed to further increase the
transition region, and establish larger free energy barrier between
the two phases. Experimentally, above $460\,$K the $\alpha-\gamma$
phase transition ends with the finite temperature critical point. Our
calculation at high temperature $900\,$K shows that the signature of
the phase transition in $F(V)$ and $E(V)$ disappears, which is
different than predicted by Gutzwiller method~\cite{Lanata:2014},
where the largest free energy barrier is found at these elevated
temperatures, but qualitatively consistent with
Ref.~\onlinecite{mcmahan2001} .

In summary, we successfully implemented the stationary formula for the
free energy of DFT+DMFT method. On the example of SrVO$_3$, FeO and Ce
metal we demonstrated that the method successfully predicts lattice
volumes in correlated solids, which are difficult for standard DFT
functionals. We also resolved controversy in the mechanism of the
$\alpha$-$\gamma$ transition in Cerium.

This work was supported Simons foundation under project "Many
Electron Problem'', and by NSF-DMR 1405303. T.B. was supported
by the Rutgers Center for Materials Theory. This research used resources of the Oak Ridge Leadership Computing Facility at the Oak Ridge National Laboratory, which is supported by the Office of Science of the US Department of Energy under Contract No. DE-AC05-00OR22725.

\bibliography{freee.bib}

\newpage
\begin{widetext}
\section{Supplementary Information: Free energy from stationary implementation of the DFT+DMFT functional}
\end{widetext}

\section{Technical Description of Computational Details}

We used the implementation of LDA+DMFT of
Ref.\onlinecite{hauleLDADMFT}, which is based on Wien2K
package~\cite{wien2k}.  The exchange-correlation functional $E_{xc}$
of LDA was utilized, in combination with nominal
double-counting (DC)~\cite{hauleLDADMFT,covalency}, which was shown to be closest to the
exact form of DC~\cite{exactDC}. We checked (on the example of Cerium)
that the exact-DC gives very similar free energy, as expected for a
stationary functional. The convergence of LDA+DMFT results is much
faster using nominal DC, hence most of results in this publication are
obtained by this simplification.

The impurity model is solved
using the hybridization expansion version of the numerically exact
continuous time QMC method~\cite{hauleCTQMC,Werner}.  Of the order of
300 LDA and 30 DMFT iterations were required for precision of $1\,$meV
per formula unit, and between $10^9-10^{10}$ Monte Carlo moves were
accepted per impurity iteration for precise enough impurity
solution. The resources of Titan supercomputer were used.

To construct the projector, the atomic-like local orbitals are used
$\braket{\vr|\phi_{lm}}=\frac{u_l(r)}{r} Y_{lm}(\hat{\vr})$. The
radial part of the local orbital $u_l(r)$ is the solution of the
scalar relativistic Dirac equation inside the muffin-tin sphere,
linearized at the Fermi level. The muffin-tin spheres are set to touch
at the lowest volume. We tested a few other forms of the projectors
defined in Ref.~\onlinecite{hauleLDADMFT}. The stationary $F(V)$ is
quite insensitive to the precise choice of projector, however, $E(V)$
changes much more.

For SrVO$_3$ calculations, we treated dynamically all five $3d$ orbitals of
Vanadium. The muffin-thin radius of Vanadium was set to $R_{mt}=1.83\,
a_B$, and $U=10\,$eV was used, which was previously shown to give good
spectra ~\cite{covalency,exactDC} for this localized orbital (see
spectra below). The
Yukawa form of screening interaction than gives $J\approx 1\,$eV (see
note below).
Brillouin zone integrations were done over $15\times 15\times 15$
k-point in the whole zone in the self-consistent calculations, and for
calculation of the impurity entropy, the hybridization is computed on
more precise $36\times 36\times 36$ k-points mesh. We mention in
passing that impurity entropy is very sensitive to the precise
frequency dependence of the hybridization, and requires very dense
momentum mesh.  

For FeO, all five $3d$ orbitals are treated by DMFT and the muffin-thin
radius of iron is set to $R_{mt}=2.11\, a_B$, and the Coulomb
repulsion to previously determined $U=8eV$~\cite{cohen2012}, which
requires $J\approx 1eV$ in Yukawa form.  In Ce metal, all seven $4f$
orbitals are treated by DMFT and the muffin-thin sphere is
$R_{mt}=2.5\,a_B$, the k-point mesh is $21\times 21\times 21$, and the
Coulomb $U=6\,$eV~\cite{McMahanU,amadon,amadon2014}, leads to
$J=0.72\,$eV in Yukawa form. The spin-orbit coupling is included in
Cerium, but neglected in SrVO$_3$ and FeO.

\section{Details on Evaluation of LDA+DMFT functional}

Here we explain how we evaluate the total energy Eq.1 and the free energy
Eq.4 of the main text.

For total energy Eq.1, we group the terms in the following way
\begin{widetext}
\begin{eqnarray}
E = \Tr((-\nabla^2+V_{ext}+V_{H}+V_{xc}) G)-\Tr((V_H+V_{xc})\rho) + E^H[\rho]+E^{xc}[\rho]+E_{nuc-nuc}+
\frac{1}{2}\Tr(\Sigma G)-\Phi^{DC}[\rho_{loc}] 
\label{Eq:Ene1}
\end{eqnarray}
\end{widetext}
We then split the energy into three terms $E=E_1+E_2+E_3$, where the first
two parts $E_1$, $E_2$ are computed using the
Green's function of the solid, and the third $E_3$ using the impurity
Green's function.

The first five terms in Eq.~\ref{Eq:Ene1} look similar to the standard
DFT energy functional, except that the Green's function $G$ here is
the self-consistent LDA+DMFT Green's function. We first solve the
eigenvalue problem for Kohn-Sham states
$(-\nabla^2+V_{ext}+V_H+V_{xc})\psi_{ik}=\varepsilon^{DFT}_{ik}\psi_{ik}$,
where $\varepsilon_{ik}^{DFT}$ are DFT-like energies, computed on
LDA+DMFT charge. We then evaluate
\begin{eqnarray}
E_1=\Tr(\varepsilon^{DFT} G) 
\end{eqnarray}
and
\begin{eqnarray}
E_2 = -\Tr((V_H+V_{xc})\rho) + E^H[\rho]+E^{xc}[\rho]+E_{nu-nuc}.
\label{Eq:E2}
\end{eqnarray}
Both $E_1$ and $E_2$ are computed using Green's function $G$ and
density $\rho$ of the solid in the same way as the standard DFT total
energy is implemented~\cite{FPLAPWTotalEnergy}.

The last two terms of Eq.~\ref{Eq:Ene1} can be computed either from the local Green's
function $\hat{P} G$ or from the impurity Green's function $G_{imp}$. Once the
self-consistency is reached, the two are of course equal. We choose to
evaluate the second term on the impurity $G_{imp}$
\begin{eqnarray}
E_3 =  \frac{1}{2}\Tr(\Sigma_{imp} G_{imp})-\Phi^{DC}[\rho_{imp}] 
\label{Eq:E3}
\end{eqnarray}
However, we never actually use Migdal-Galitskii formula,
because it is numerically much less stable than computing the
potential energy from the impurity probabilities, i.e.,
$$\frac{1}{2}\Tr(\Sigma_{imp} G_{imp})=\sum_m P_m E^{atom}_m - \Tr(\varepsilon_{imp} n_{imp})$$

The free energy functional $\Gamma[G]$ (Eq.~2 of the main text) is
\begin{widetext}
\begin{eqnarray}
\Gamma[G] = \Tr\log G-\Tr\log((G_0^{-1}-G^{-1})G) + E^{H}[\rho]+E^{xc}[\rho]+\Phi^{DMFT}[G_{loc}]-\Phi^{DC}[\rho_{loc}]+E_{nuc-nuc}.
\label{Eq:Funcq}
\end{eqnarray}
\end{widetext}
First, we extremize it ($\delta\Gamma[G]/\delta G=0$) to obtain the Dyson equation  
\begin{eqnarray}
G^{-1}-G_0^{-1} + V_H + V_{xc}+\Sigma_{DMFT}-V_{dc}=0.
\end{eqnarray}
A note is in order here. We assumed $\delta P/\delta G=0$, which holds
whenever the projector does not depend on the self-consistent charge
density. To ensure this property, we used for the localized orbitals
$\ket{\phi}=\frac{u_l(r)}{r}Y_{lm}(\hat{r})$, where the radial wave function
$u_l(r)$ is the solution of the scalar relativistic Dirac equation on
the LDA charge density (rather than self-consistent charge
density). Note also that the use of the self-consistently determined
Wannier functions (which depend on self-consistent charge), as is
commonly used in most of the LDA+DMFT
implementations~\cite{park2014a,park2014b,lechermann2012}, leads to
non-stationary LDA+DMFT solution, and non-stationary free energies.

We next insert the expression $G^{-1}-G_0^{-1}$ into
Eq.~\ref{Eq:Funcq} to obtain expression for free energy
\begin{eqnarray}
F = 
E_{nuc-nuc}-\Tr((V_H + V_{xc})\rho)+
E^{H}[\rho]+E^{xc}[\rho]\nonumber\\
+\Tr\log G
-\Tr(\Sigma_{DMFT} G) 
+ \Phi^{DMFT}[G_{loc}]\nonumber\\
+\Tr(V_{dc}\rho_{loc}) 
-\Phi^{DC}[\rho_{loc}]+\mu N
\end{eqnarray}
The impurity free energy $F_{imp}$ contains $\Phi^{DMFT}[G_{imp}]$ in
the following way
\begin{eqnarray}
F_{imp} = \Tr\log G_{imp} - \Tr(\Sigma_{imp} G_{imp}) + \Phi^{DMFT}[G_{imp}].
\end{eqnarray}
In DMFT, $G_{loc}=G_{imp}$ and $\Sigma_{DMFT}=\Sigma_{imp}$, hence 
we can write
\begin{eqnarray}
F = 
E_{nuc-nuc}-\Tr((V_H + V_{xc})\rho)+E^{H}[\rho]+E^{xc}[\rho]\nonumber\\
+ \Tr\log(G)-\Tr\log(G_{loc})+F_{imp}\nonumber\\
+\Tr(V_{dc} \rho_{loc}) -\Phi^{DC}[\rho_{loc}]+\mu N.
\end{eqnarray}
This equation appears as Eq.4 in the main text.

Next we split free energy of the impurity into the energy and the entropy term
$$F_{imp} = E_{imp}-T S_{imp},$$
where
\begin{eqnarray}
E_{imp}=
\Tr((\Delta+\varepsilon_{imp}-\omega_n\frac{d\Delta}{d\omega_n})G_{imp}) 
\nonumber\\
+ \frac{1}{2}\Tr(\Sigma_{imp} G_{imp}) -T S_{imp}
\end{eqnarray}
Hence
\begin{eqnarray}
F=
\frac{1}{2}\Tr(\Sigma_{imp} G_{imp})-\Phi^{DC}[\rho_{loc}]
-T S_{imp} 
\nonumber\\
+E_{nuc-nuc}-\Tr((V_H + V_{xc})\rho)+
E^{H}[\rho]+E^{xc}[\rho]\nonumber  \\
+\Tr\log(G)-\Tr\log(G_{loc})
+\Tr(V_{dc} \rho_{loc})+\mu N \nonumber\\
+\Tr((\Delta+\varepsilon_{imp}-\omega_n\frac{d\Delta}{d\omega_n})G_{imp}) 
\end{eqnarray}
Again using the identity $G_{imp}=G_{loc}$ and $\rho_{imp}=\rho_{loc}$
as well as the definition of $E_2$ (Eq.~\ref{Eq:E2}) and $E_3$ (Eq.~\ref{Eq:E3}) we obtain
\begin{eqnarray}
F&=& \Tr\log(G)+\mu N+E_2\nonumber\\
&+&\Tr((\Delta -\omega_n\frac{d\Delta}{d\omega_n}+\varepsilon_{imp}+V_{dc})G_{loc}) \nonumber\\
&-&\Tr\log(G_{loc})+E_3-T S_{imp}
\label{Eq:Fn}
\end{eqnarray}
This equation is implemented in our DFT+DMFT code.
Similarly than in the implementation of the total energy Eq.~\ref{Eq:Ene1},
we compute $E_3$ and $T S_{imp}$ using impurity quantities, while the
rest of the terms are computed using the Green's function of the
solid. [Alternatively, we could also compute the last two rows using
impurity quantities, and the first row using the solid Green's function].
In this way we ensure that $F$ and $E$ are split in the same way
between the "impurity'' and the "lattice'' quantities, hence they
share almost identical Monte Carlo noise. However, when comparing
$E(V)$ and $F(V)$ at two different volumes, $F(V)$ converges faster
that $E(V)$ with the number of LDA and/or DMFT iterations.
Moreover, $F(V)$ is very robust with respect to small changes in
projector or double-counting, while $E$ is more sensitive.

Notice that $F+T S_{imp}$ can be evaluated at each LDA+DMFT iteration,
just like the total energy above. To add $T S_{imp}$ at low
temperatures, we however need a few extra impurity runs. The method of
computing $T S_{imp}$ is explained in the main text, and requires the
impurity energy at a few temperatures. An alternative to this approach
is to compute $T S_{imp}$ from so called "flat-histogram sampling
method"~\cite{Gull}, which is also done as postprocessing on
self-consistent LDA+DMFT hybridization $\Delta$.

Perhaps, the most challenging term in Eq.~\ref{Eq:Fn} to compute is
$\Tr\log(G)$, which requires eigenvalues (but not eigenvectors) of the
LDA+DMFT eigenvalue problem. We first diagonalize
\begin{eqnarray}
(-\nabla^2+V_{ext}+V_H+V_{xc}+\Sigma(i\omega_n)-V_{dc})\psi_{i,k,\omega_n}=\nonumber\\
=\varepsilon_{i,k,\omega_n}\psi_{i,k,\omega_n}.
\end{eqnarray}
and then evaluate
\begin{widetext}
\begin{eqnarray}
\Tr\log(G)+\mu N=
T\sum_{i\omega_n,i,k,\sigma}\left(\log(\varepsilon_{i,k,\omega_n}-i\omega_n-\mu)-\log(\varepsilon_{i,k,\infty}-i\omega_n-\mu)\right)
-T\sum_{i,k,\sigma}\log(1+e^{-\beta(\varepsilon_{i,k,\infty}-\mu)})+\mu N
\end{eqnarray}
\end{widetext}
Here it becomes apparent that if $\Sigma(i\omega_n)$ is
frequency independent, the first term in the brackets vanishes,
while the second term gives (at $T=0$) the sum of eigenvalues $$\Tr\log(G)+\mu N
\rightarrow^{U=0}\rightarrow \sum_{i,k,\sigma}\theta(\varepsilon_{i,k}<\mu)\,\varepsilon_{i,k},$$
the well known DFT contribution to the total energy.

\section{Comparison with standard functionals}

Here we compare total energy of LDA, PBE~\cite{PBE}, and PBEsol~\cite{PBEsol} functionals with
the free energy of LDA+DMFT.

In most weakly correlated solids, LDA underestimates lattice constants
on average for 1.6\%, while PBE~\cite{PBE} overestimates them for
approximately 1\%.~\cite{Kresse} PBEsol~\cite{PBEsol} was designed to
predict most accurate volumes in solids, and it typically falls
in-between LDA and PBE.

\begin{figure}
\includegraphics[width=0.86\linewidth]{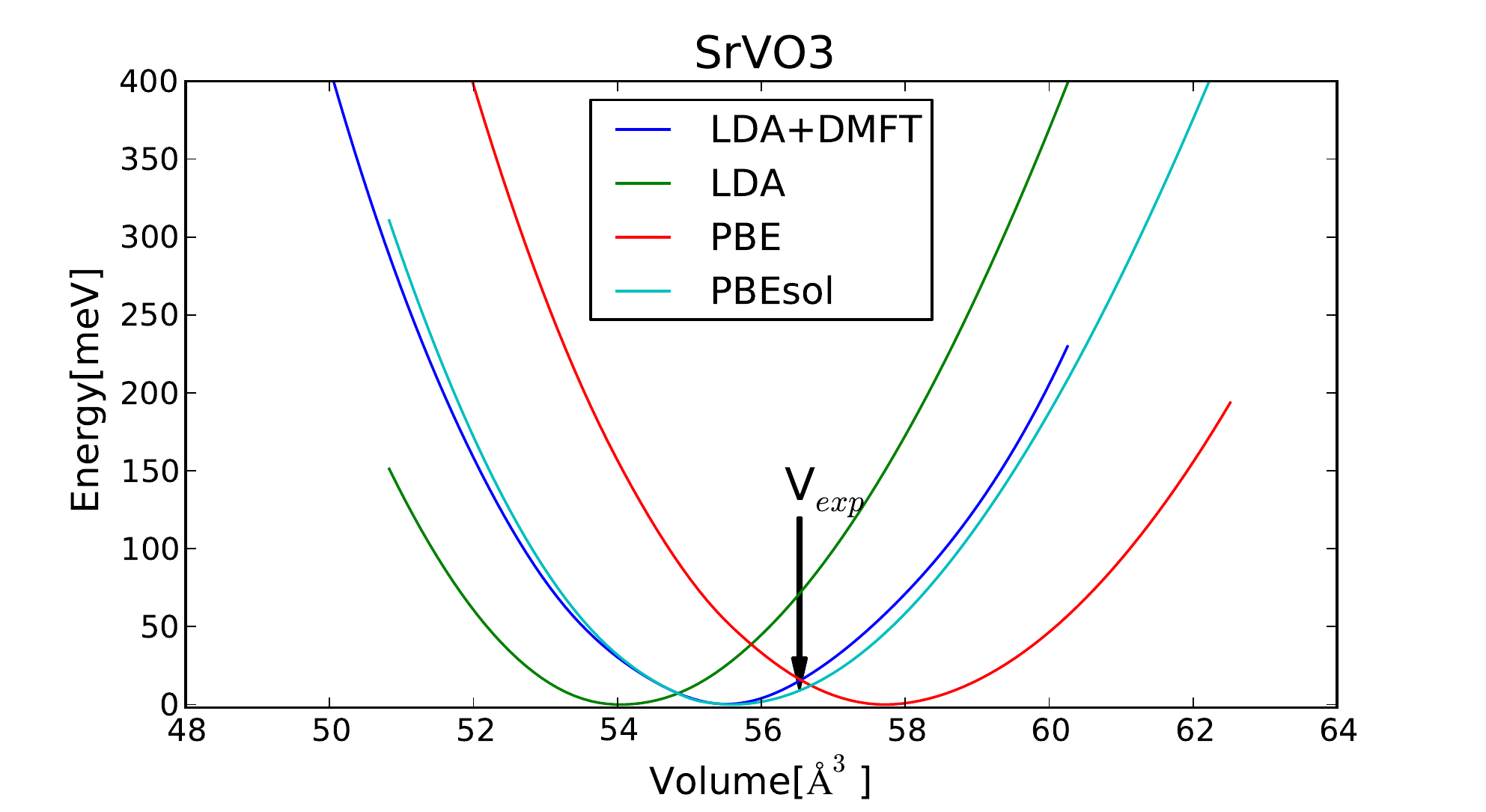}
\caption{
Free energy of LDA+DMFT for SrVO$_3$ compared with total energy of other standard
DFT functionals.
}
\label{fig:SrVO3f}
\end{figure}
In Fig.~\ref{fig:SrVO3f} we compare LDA+DMFT free energy in SrVO$_3$ with the
total energy computed by other functionals. Both LDA+DMFT and
PBEsol underestimate lattice constant for approximately 0.6\%, while
LDA underestimates it for  1.5\%, and PBE overestimates for
0.7\%. Hence predictions of standard functionals in the case of
SrVO$_3$ are quite in line with standard performance in weakly
correlated solids. Perhaps, this is not very surprising given that
SrVO$_3$ is a metallic moderately correlated system.

\begin{figure}
\includegraphics[width=0.86\linewidth]{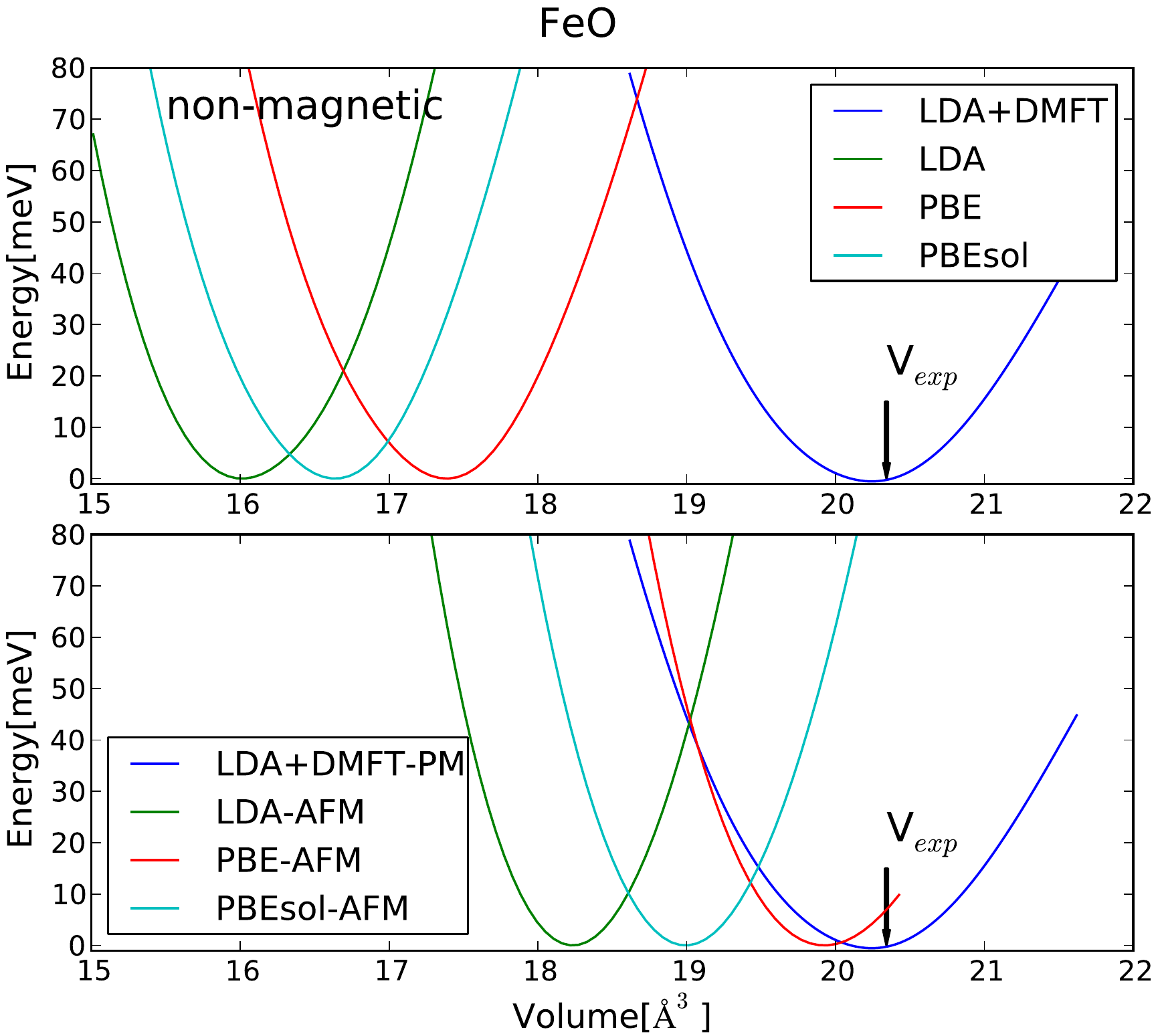}
\caption{
Free energy of LDA+DMFT for FeO compared with total energy of other standard
DFT functionals. Upper (lower) panel shows non-magnetic
(antiferromagnetic) DFT calculation. LDA+DMFT results are obtained at
300K in paramagnetic state.
}
\label{fig:FeOf}
\end{figure}

In FeO (Fig.~\ref{fig:FeOf}), all standard functionals severally
underestimate volume in the paramagnetic state. For example the
lattice constants with LDA, PBEsol and PBE are 7.7\%, 6.5\% and 5.1\%
too small, far outside the standard performance of these functionals
in weakly correlated solids.

The predictions are
improved when the AFM long range order is allowed. LDA and PBEsol still
underestimate lattice constant for  3.6\%, and 2.3\% respectively. On
the other hand PBE is this time quite close to the experiment
(underestimates for 0.7\%). In comparison LDA+DMFT underestimates it
for only 0.16\%. It is quite clear that the excellent prediction of
AFM-PBE here is
merely a coincidence, as normally PBE overestimates the volume.

\begin{figure}
\includegraphics[width=0.86\linewidth]{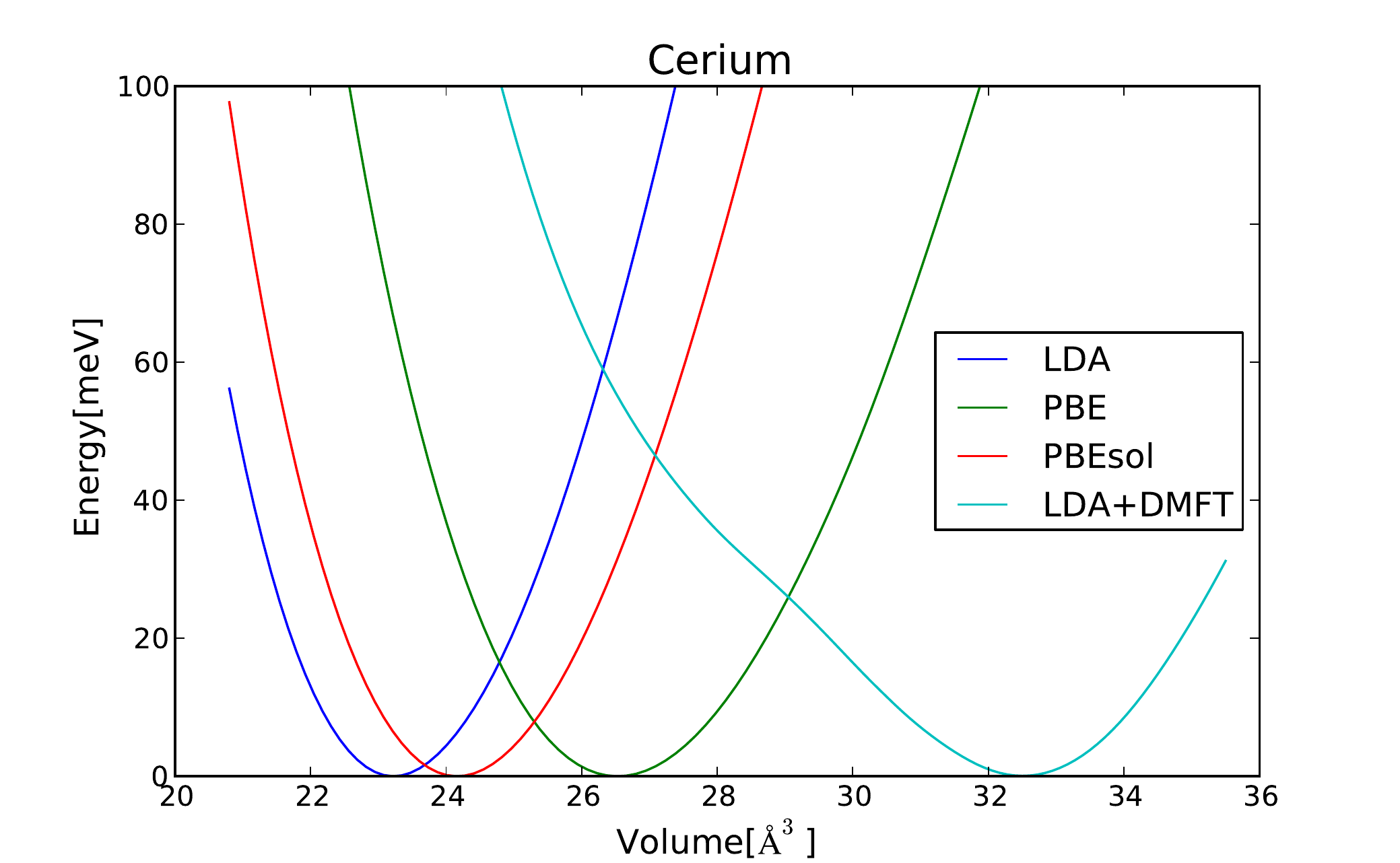}
\caption{
Free energy of LDA+DMFT for Cerium compared with total energy of other 
DFT functionals. LDA+DMFT results are obtained at $400\,$K.
}
\label{fig:Cef}
\end{figure}

Finally, we plot results for Cerium in Fig.~\ref{fig:Cef}. The result
of LDA+DMFT is very different from those of any other functional, as
it clearly contains the nontrivial soft mode for the
$\alpha$-$\gamma$ transition. No other functional shows any hint of
such transition.

The equilibrium volume in Cerium is strongly temperature dependent,
and is approximately $34\Ang^3$ at zero pressure and $400\,$K, while it
changes to approximately $28\Ang^3$ in the $\alpha$ phase at low
temperature. The LDA+DMFT results are computed at $400\,$K, hence at
$p=0$ the volume is somewhat underestimated (1.5\%), but under
pressure (already at $1\,$GPa) the agreement with experiment
is considerably improved. 

The DFT results should be compared to $T=0$ experimental volume  of
28\,$\Ang^3$. All functionals underestimate the lattice constant, LDA
for $6\%$, PBEsol for $5\%$ and PBE for $1.8\%$. Clearly electronic
correlations are very important  even in the $\alpha$ phase at low
temperature, as standard DFT functionals substantially underestimate
the volume.

\section{Screened Coulomb repulsion of Yukawa form}

It is noted above that we used the Yukawa representation of the
screened Coulomb interaction, in which there is unique relationship
between the Hubbard $U$ and Hund's coupling $J$. If $U$ is specified,
$J$ is uniquely determined. To show this we derive the matrix elements
of screened Coulomb interaction in our DMFT orbital basis 
\begin{widetext}
\begin{eqnarray}
U_{m_1 m_2 m_3 m_4}=\int d^3 r\int d^3 r' 
\left(\frac{u_l(r)}{r}\right)^2 \left(\frac{u_l(r')}{r'}\right)^2 
Y_{l m_1}^*(\hat{\vr}) Y_{l m_4}(\hat{\vr}) Y_{l m_2}^*(\hat{\vr}') Y_{l m_3}(\hat{\vr}')
\frac{e^{-\lambda |\vr-\vr'|}}{|\vr-\vr'|}
\label{Eq:U}
\end{eqnarray}
There exist a well known expansion of Yukawa interaction in terms of spheric harmonics $Y_{km}$, which reads
\begin{eqnarray}
\frac{e^{-\lambda|\vr-\vr'|}}{|\vr-\vr'|} = 4 \pi\sum_{k} \frac{I_{k+1/2}(r_<) K_{k+1/2}(r_>)}{\sqrt{r_<\; r_>}}\sum_m Y_{km}^*(\hat{\vr}) Y_{km}(\hat{\vr}')
\end{eqnarray}
Here $r_<=min(r,r')$, $r_>=max(r,r')$, $I$ and $K$ are modified Bessel function of the first and second kind.
Inserting this expression into Eq.~\ref{Eq:U}, we get
\begin{eqnarray}
U_{m_1 m_2 m_3 m_4}=\sum_k \frac{4\pi}{2k+1}\braket{Y_{l m_1}|Y_{k m_1-m_4}|Y_{l m_4}}
\braket{Y_{l m_2}|Y^*_{k m_3-m_2}|Y_{l m_3}}\nonumber\\
\times (2k+1)\int_0^\infty dr \int_0^\infty dr' u_l^2(r) u_l^2(r')  \frac{I_{k+1/2}(\lambda r_<) K_{k+1/2}(\lambda r_>)}{\sqrt{r_<\; r_>}}.
\end{eqnarray}
Hence, the screened Coulomb interaction has the Slater form with the Slater integrals being
\begin{eqnarray}
F^k = (2k+1)\int_0^\infty dr \int_0^\infty dr' u_l^2(r) u_l^2(r')  \frac{I_{k+1/2}(\lambda r_<) K_{k+1/2}(\lambda r_>)}{\sqrt{r_<\; r_>}}.
\label{Eq:Slater}
\end{eqnarray}
\end{widetext} 
This is a product of two one-dimensional integrals and is very easy to efficiently implement.

It is clear from Eq.~\ref{Eq:Slater} that $\lambda$ uniquely determines all $F^k$'s, and furthermore even one Slater integral ($F^0$) uniquely determines $\lambda$. This is because $F^k$ are monotonic functions of $\lambda$ and take the value of bare $F^k$ at $\lambda=0$ and vanish at large $\lambda$. 
Hence given $F^0$, the screening length $\lambda$ is uniquely determined, and hence other higher order $F^k$ are uniquely determined as well.

\section{Mass renormalization of metallic $SrVO_3$}

\begin{figure}
\includegraphics[width=0.9\linewidth]{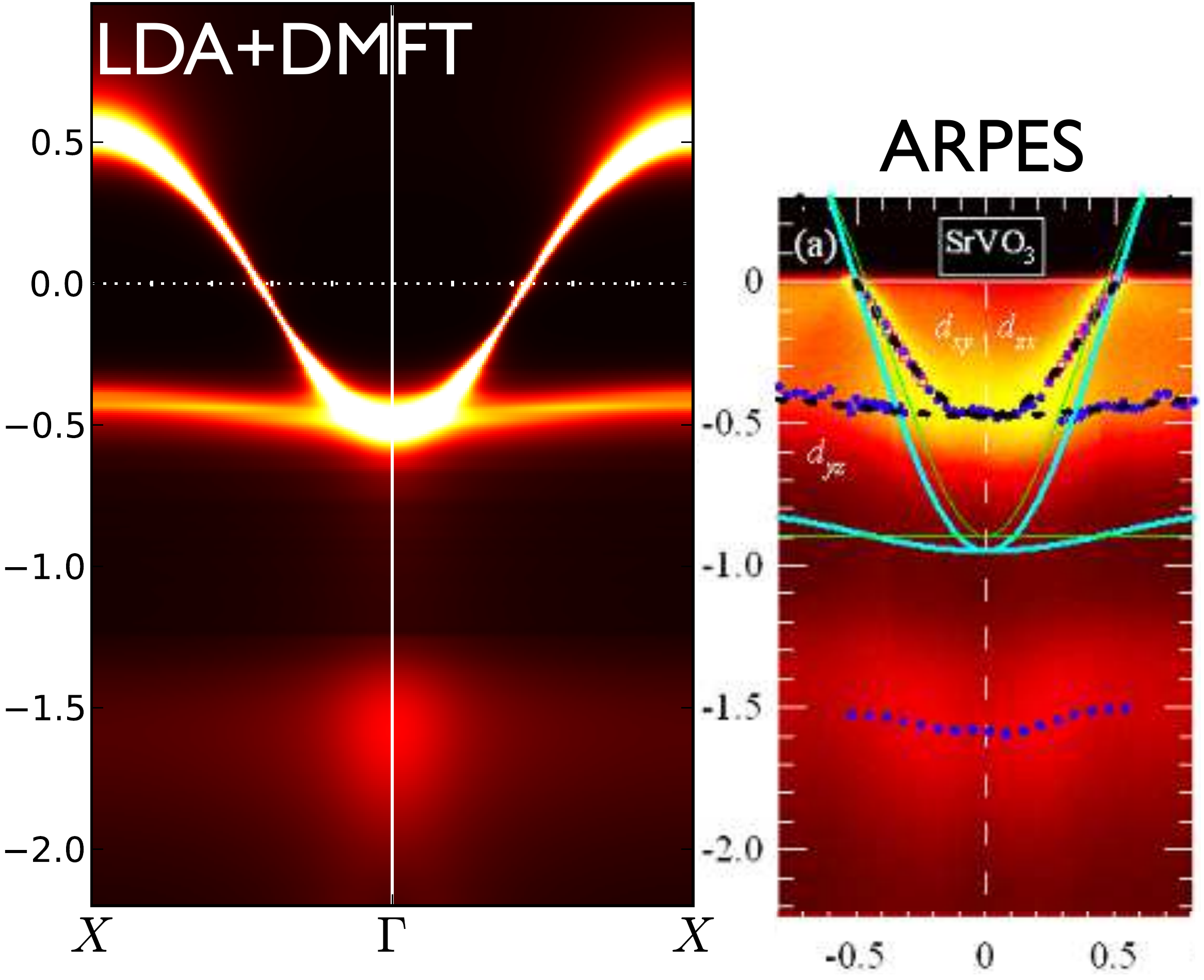}
\caption{
Spectral function of SrVO$_3$ within LDA+DMFT at equilibrium volume
compared with ARPES spectra of Ref.~\onlinecite{SVO_ARPES}.
}
\label{fig:SVO_ARPES}
\end{figure}

Even though the Coulomb interaction in SrVO$_3$ is $U=10\,$eV,  it
gives a relatively moderate mass enhancement over DFT band structure in
all-electron LDA+DMFT implementation. This is because the interaction
is severely screened by hybridization of $d$ states with oxygen $p$
states, and because the $t_{2g}$ orbitals are in mixed-valence state
($n_{t2g}\approx 1.5$)~\cite{exactDC,covalency}.
In Fig.~\ref{fig:SVO_ARPES} we show the LDA+DMFT spectral function as
well as recent APRES measurements~\cite{SVO_ARPES}. The mass
renormalization in the $t_{2g}$ orbital is $m^*_{t2g}/m_{band}\approx 2$ 
and in $e_g$ is $m^*_{t2g}/m_{band}\approx 1.3$
The agreement between ARPES spectra (the experimental signal is color
coded on the right) and LDA+DMFT spetrcal fuction $A(k,\omega)$
(plotted on the left) is very good, both in the quasiparticle band
(between $-0.5\,$eV and $0.5\,$eV) and Hubbard sattelite at $-1.5\,$eV.

\end{document}